%% file: tls.tex
\documentclass[aps,prb,reprint,twoside,twocolumn,superscriptaddress,showpacs,showkeys,footinbib]{revtex4-1}
\usepackage{amsmath}
\usepackage{graphicx}
\usepackage{verbatim}
\usepackage{color}
\usepackage{txfonts} 
\usepackage{float} 
\usepackage{subfigure}
\usepackage[bookmarks=true,colorlinks=true,urlcolor=blue,linkcolor=blue,citecolor=blue]{hyperref}

\newcommand{\bea}{\begin{eqnarray}}
\newcommand{\eea}{\end{eqnarray}}

\newcommand{\ai}{\textit{ab initio}}
\newcommand{\rp}{\right)}
\newcommand{\lp}{\left(}
\newcommand{\rc}{\right]}
\newcommand{\lc}{\left[}

\newcommand{\ra}{\right\rangle}
\newcommand{\la}{\left\langle}

\newcommand{\bfa}{(a)}
\newcommand{\bfb}{(b)}
\newcommand{\bfc}{(c)}

\def\myeq#1{Eq.~(\ref{#1})}
\def\mytab#1{Table~\ref{#1}}
\def\myfig#1{Fig.~\ref{#1}}
\def\mysec#1{Sec.~\ref{#1}}

\begin{document}

\title{Identification of structural motifs as tunneling two-level systems in amorphous alumina at low temperatures}

\author{Alejandro P\'{e}rez Paz}
\affiliation{Nano-Bio Spectroscopy Group and ETSF Scientific Development Centre, Departamento de F\'{i}sica de Materiales, Centro de F\'{i}sica de Materiales CSIC-UPV/EHU-MPC and DIPC, Universidad del Pa\'{i}s Vasco UPV/EHU, E-20018 San Sebasti\'{a}n, Spain}

\author{Irina V. Lebedeva}
\affiliation{Nano-Bio Spectroscopy Group and ETSF Scientific Development Centre, Departamento de F\'{i}sica de Materiales, Centro de F\'{i}sica de Materiales CSIC-UPV/EHU-MPC and DIPC, Universidad del Pa\'{i}s Vasco UPV/EHU, E-20018 San Sebasti\'{a}n, Spain}

\author{Ilya V. Tokatly}
\affiliation{Nano-Bio Spectroscopy Group and ETSF Scientific Development Centre, Departamento de F\'{i}sica de Materiales, Centro de F\'{i}sica de Materiales CSIC-UPV/EHU-MPC and DIPC, Universidad del Pa\'{i}s Vasco UPV/EHU, E-20018 San Sebasti\'{a}n, Spain}
\affiliation{IKERBASQUE, Basque Foundation for Science, E-48011 Bilbao, Spain}

\author{Angel Rubio}
\email{angel.rubio@ehu.es}
\affiliation{Nano-Bio Spectroscopy Group and ETSF Scientific Development Centre, Departamento de F\'{i}sica de Materiales, Centro de F\'{i}sica de Materiales CSIC-UPV/EHU-MPC and DIPC, Universidad del Pa\'{i}s Vasco UPV/EHU, E-20018 San Sebasti\'{a}n, Spain}

\begin{abstract}
One of the most accepted models that describe the anomalous thermal behavior of amorphous materials at temperatures below 1~K relies on the quantum mechanical tunneling of atoms between two nearly equivalent potential energy wells forming a two-level system (TLS).
Indirect evidence for TLSs is widely available.
However, the atomistic structure of these TLSs remains an unsolved topic in the physics of amorphous materials.
Here, using classical molecular dynamics, we found several hitherto unknown bistable structural motifs that may be
key to understanding the anomalous thermal properties of amorphous alumina at low temperatures.
We show through free energy profiles that the complex potential energy surface can be reduced to canonical TLSs. 
The predicted tunnel splittings from instanton theory, the number density, dipole moment, and coupling to external strain of the discovered motifs are consistent with experiments.
\end{abstract}

\maketitle

\section{Introduction}

Since the first measurements of the thermal behavior of glasses at temperatures below 1~K, it became clear that such systems possess very peculiar properties that unite almost all disordered materials regardless their chemical composition and bonding~\cite{phillips81, phillips87, pohl02, klinger10} and distinguish them from their crystalline counterparts.
Namely, the heat capacity and thermal conductivity of amorphous systems increase with temperature almost linearly and quadratically, respectively, which are in sharp contrast to the Debye T$^3$ behavior of crystals~\cite{NATO01}. 
This anomalous behavior was attributed to microscopic tunneling two-level systems (TLSs) that are sparsely present in disordered materials~\cite{phillips72, anderson72}.

The same TLSs are believed to be the main source of decoherence and noise in Josephson junction-based superconducting qubits ~\cite{simmonds04, martinis05}, various nanomechanical~\cite{zolfagharkhani05, hoehne10} and optical~\cite{arcizet09} resonators using amorphous materials. Exacerbating the problem is that each sample has its own set of TLS frequencies that cannot be controlled at will~\cite{simmonds04, lupascu09, lisenfeld10a, lisenfeld10b, palomaki10}. 
 As often happens, the phenomenon that is parasite for many applications at the same time opens the way to others. Such applications, as TLS-based qubits~\cite{sun10, grabovskij11}, devices for non-linear optical measurements~\cite{ramos13} or quantum memory~\cite{zagoskin06} based on TLSs, have been proposed.  
However, inability to control TLSs and lack of understanding of their atomistic structure hinders these developments as well. 

The recent breakthrough experiments by Grabovskij \textit{et al}.~\cite{grabovskij12} on microwave response of individual TLSs under strain leave no doubt that at least in the Josephson junctions with the amorphous alumina (Al$_2$O$_3$) barrier, the low energy excitations are described by the standard TLS model. Nevertheless, no consensus has been reached yet on the nature of the tunneling particles. 
The fact they should carry a substantial electric charge 
that couples to the Josephson junction electric field has generated interpretations based on tunneling of electrons~\cite{agarwal13, sousa09, faoro06, lutchyn08}. 
To fit into the experimental range of small energy splittings, additional arguments had to be invoked such as polaronic dressing of tunneling electrons~\cite{agarwal13}.  
TLS models based on the delocalization of a single oxygen atom~\cite{dubois13} or hydrogen impurities~\cite{khalil13, holder13} have been also proposed. 
We think, however, that electron-based models may not be needed because atoms in an ionic material such as amorphous alumina already carry significant partial charges.
~\footnote{Of course, the co-existence of both intrinsic and extrinsic TLSs cannot be excluded in real amorphous materials.}
Neither is there the need to assume the presence of any specific impurities.
The disordered alumina structure should already contain intrinsic bistable motifs.
Here, we undertake the computational search and characterization for such structural motifs. 

The paper is organized as follows. 
In \mysec{tls}, we review the standard TLS model. 
The computational methods and details are described in \mysec{methods}
In \mysec{discussion}, we present and discuss the structural motifs found in our computational simulations on amorphous alumina at low temperatures.
Finally, the conclusions are drawn in \mysec{conclusions}.

\section{The TLS model}
\label{tls}

In 1972, Phillips~\cite{phillips72} and Anderson \textit{et al}.~\cite{anderson72} proposed independently that in amorphous materials some atoms could tunnel quantum-mechanically between two nearly isoenergetic configurations forming TLS, just like the umbrella motion of the ammonia molecule.
They assumed that this scenario could be modelled by an effective double well potential whose barrier height $V_0$ and energy difference between the two minima $\Delta$ were broadly distributed across the amorphous sample.
The energy splitting between the two lowest vibrational eigenstates should be small ($\sim$~0.1~meV) to contribute to the properties of glasses at temperatures below 1~K~\cite{phillips81, phillips87,  martinis05, simmonds04, lupascu09, lisenfeld10a, lisenfeld10b, palomaki10, grabovskij12}.
This tunneling splitting is given by $\delta E=\sqrt{\Delta^2 + \Delta_0^2}$,
where the coupling energy $\Delta_0=\hbar\Omega\exp{\lp-d\sqrt{2 m_\mathrm{eff} V_0}/\hbar\rp}$ is proportional to the overlap between the unperturbed 
localized wavefunctions in each well~\cite{phillips81, phillips87, klinger10}.
Here, $d$ is the well separation, $\Omega$ is a typical well vibration frequency, and $m_\mathrm{eff}$ is the effective mass of the tunneling particle. 

The energy asymmetry between the two wells of a TLS can be also varied using an external strain field as $\delta{\Delta}=2 \gamma \epsilon$, where $\epsilon$ is the dimensionless strain and $\gamma$ is the strain-asymmetry coupling that was measured to be about 1~eV~\cite{grabovskij12, heuer94, phillips81, phillips87}. 
The coupling between the TLSs and an external electric field $\mathbf{F}$ is of the dipole type $\delta{\Delta}=2\mathbf{p\cdot F}$~\cite{simmonds04, martinis05, lupascu09, lisenfeld10a, lisenfeld10b, palomaki10, grabovskij12} with the TLS dipole moment $\mathbf{p}$  corresponding to a single effective charge moving by a distance of a single atomic bond~\cite{martinis05, constantin07}.

These TLSs are sparsely present in the bulk of amorphous materials ($\sim 10^{21}$~eV$^{-1}$~cm$^{-3}$ according to experiments~\cite{grabovskij12, phillips81, phillips87}) so measurements on those samples yield averages over many different TLSs. 
This explains why the experimental verification of the TLS model remains extremely difficult and, though supported by numerous indirect data~\cite{phillips81, phillips87,  klinger10}, is still hotly debated~\cite{leggett91, leggett13}.  
Modern extensions of the original TLS model, including the ``soft-mode'' model~\cite{klinger80} and the ``defect-interaction'' model~\cite{springer98,leggett88}, were recently reviewed~\cite{klinger10}. 

\section{Computational Methods and Details}
\label{methods}

From the computational viewpoint, finding TLSs in amorphous materials is difficult due to the required large system sizes and long time scales. 
Firstly, the low density of TLS imposes severe limitations on the sizes of the systems that have to be involved in search for TLS. 
Secondly, the necessity of descending to very low temperatures to freeze all structural transformations except for flips between the configurations forming TLS makes these processes hardly activated. 
As a result, TLSs appear as extremely ``rare events'' both in space and time, which virtually excludes First-Principles methods from the available theoretical tools. 
For this reason, accurate and transferable force fields have been instead adopted in most computer simulations in this research field.
To add further complication, the quenching simulated by computers is typically several orders of magnitude faster than the natural cooling process~\cite{binder96}.
Moreover, there is no way to guarantee that TLSs are present after the melting-quenching computational protocol used to generate the amorphous configurations.
If present, the TLSs are metastable entities that are easily destroyed by a variety of physical factors such as temperature and/or mechanical strain.
A final caveat for computer simulations is that one should use a computational cell with a number of atoms much larger than the atoms comprising the TLS, which is often unknown \textit{a priori}.
In summary, these limitations in time scales and system sizes --and hence the need of much sampling-- render the use of \textit{ab initio} techniques prohibitive.

Nowadays, classical molecular dynamics (MD) represents the best compromise between computational cost and accuracy in the quest for TLS candidates. 
For instance, in amorphous silica (SiO$_2$), TLSs were reported to involve large cooperative re-orientations
or coupled rigid rotation of nearly 30 SiO$_4$ tetrahedra with an estimated energy barrier of 0.06~eV~\cite{buchenau84,trachenko98,trachenko00}.
The MD approach adopted in this work entails the following three consecutive steps: 
(1) Generating amorphous configurations of alumina using a standard melting-and-quenching protocol~\cite{gutierrez02}.
(2) Searching for any large amplitude fluctuation in atomic positions over time during MD simulations.
(3) Finally, for each large hop, we reconstruct the free energy profile to assess whether it qualifies as a TLS candidate.
We calculate the tunnel splittings on the potential energy profile obtained at 0~K for the successful TLS candidates.

In all calculations, we used two force fields, hereafter called, I~\cite{matsui94} and II~\cite{beck}, to check the reproducibility and robustness of our motifs. 
The first of these force-fields (I) has been extensively tested against experimental properties of crystalline~\cite{matsui96pcm, matsui96grl, belonoshko, ahuja}, liquid \cite{gutierrez00,gutierrez11} and amorphous alumina \cite{gutierrez02}.
The second force-field (II) was fitted to \textit{ab initio} data on crystalline and liquid alumina~\cite{beck}. 
The structural properties of our alumina systems have been compared to previous calculations (see \myfig{rdf}) and excellent agreement is found.
The definitions of force fields I and II are presented in \mytab{ffI} and \mytab{ffII}, respectively.
\input{./ff.tex}

Two cell sizes were considered for the MD simulations of alumina: 
a medium-sized cell containing 1500 atoms (300 Al$_2$O$_3$ units) and a smaller one with 360 atoms (72 Al$_2$O$_3$ units).
Different MD codes and simulation parameters were used for each system size.
In all cases, we cross-checked our findings with the different simulation protocols employed.

\begin{figure*}
\includegraphics[width=\textwidth]{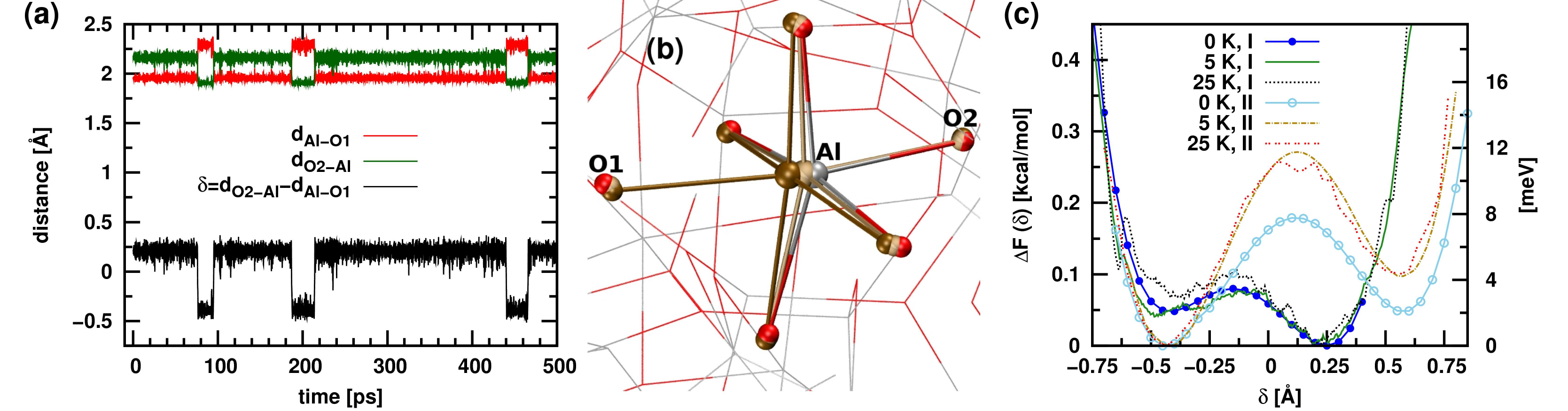}
\caption{\label{1stmotif}
(Color online) Found bistable motif: a distorted octahedron of oxygen atoms surrounding a central aluminum atom jumping between oxygen atoms O1 and O2.
\bfa\ Calculated time evolution of the relevant collective variables during a microcanonical MD simulation of 1500 alumina atoms at 5~K using the force field I.
\bfb\ Superimposed snapshots of the optimized configurations corresponding to the initial (dark ochre color), transition  (light color) and final (aluminum atoms in grey, oxygen atoms in red) states at 0~K.
At the transition state, the first coordination shell of the motif resembles a distorted octahedron with four oxygen atoms in the equatorial plane and a central aluminum atom, which oscillates between the apical oxygen atoms O1 and O2.
\bfc\ Computed free energy profiles for the aluminum transfer at 0, 5, and 25~K using force fields I~\cite{matsui94} and II~\cite{beck}.
The collective variable was the asymmetric stretch $\delta=|\vec{r}_\mathrm{O2-Al}|-|\vec{r}_\mathrm{Al-O1}|$.
}
\end{figure*}

All samples of amorphous alumina were generated following the standard quench-and-melt protocol described in Ref.~\cite{gutierrez02}.
The liquid was obtained from the melting of the experimental crystalline structure of $\alpha$-alumina (30 atoms per hexagonal cell with parameters $a=b=4.76020$~\AA, $c=12.9933$~\AA, $\alpha=\beta=90^\circ, \gamma=120^\circ$ and density of 3.98~g~cm$^{-3}$).
Liquid alumina was equilibrated at high temperature 5000~K and low density 2.75~g~cm$^{-3}$ during 200~ps.
The system was then cooled down to 3000~K at a cooling rate of 20~K~ps$^{-1}$. 
The resulting configuration was subsequently equilibrated at 3000~K for over 100~ps. 
Finally, the systems were further quenched to low temperature (25 and 5~K) at a cooling rate of 4~K~ps$^{-1}$. 
Our radial distribution functions (see \myfig{rdf}) agree very well with the previous MD simulations~\cite{gutierrez02, beck}
and demonstrate that our structures are truly amorphous.

The amorphous system with 1500 atoms equilibrated to a final cell volume of $24.1335\times 24.1335\times 26.3496$~\AA$^3$, yielding a density of 3.24~g~cm$^{-3}$. 
The electrostatics was handled with the standard Ewald summation technique.
A cutoff radius of 11~\AA\ was used to truncate the short-ranged interactions.
During equilibration, we controlled the temperature and pressure using the Nos\'e-Hoover implementations of the thermostats~\cite{nhc} and barostat~\cite{mtk}, respectively. 
A time step of 0.4~fs was used to ensure a stable integration of the equations of motion.
These calculations were performed using the freely available codes {\tt CP2K}~\cite{cp2k} and {\tt LAMMPS}~\cite{lammps}.

The molecular dynamics simulations for the 360-atoms system were performed with the {\tt MD-kMC} code~\cite{mdkmc}.
The long-range Coulomb interactions were also calculated with the standard Ewald summation technique.
The system was equilibrated to a final cubic cell of volume $15.61777^3$~\AA$^3$ after isotropic compression to reach a density of 3.20~g~cm$^{-3}.$
A cutoff of 7.81~\AA\ was used to truncate the short-ranged interactions.
The time step was 1~fs. The Berendsen thermostat was employed to control the temperature~\cite{berendsen}.

To study the dynamics of amorphous systems at low temperatures and find possible two-level systems (TLSs),
we conducted long microcanonical (constant energy) MD simulations on the equilibrated amorphous configurations for over 1~ns.
We carefully monitored the time evolution of the atomic fluctuations about their average positions. 
We selected those atoms that deviate from their average positions more than 0.15~\AA\ during time intervals longer than 1~ps. 
To search for such fluctuations, in practice we ran up to a hundred trajectories, which is statistically more efficient than performing long simulations on larger systems.
While at temperature above 50~K many of such fluctuations happen simultaneously and render the analysis complicated,
at temperatures below 10~K they are too rare and not accessible by standard MD simulations.
In practice, we chose an intermediate temperature ($\sim$~25~K) to search for bistable motifs.
After visual inspection of the MD trajectories, a collective variable $\delta$ (typically an asymmetric stretch) was chosen to describe the observed transformations.
All visualization and snapshots were done with the {\tt VMD} code~\cite{vmd}.

Free energy profiles were computed for the found ``rare events'' to check whether they qualify as TLSs, that is, to confirm their bistability.
To this end, we used and compared a variety of well-established techniques, including unbiased long MD runs, the metadynamics method of Laio and Parrinello~\cite{mtd}, 
the average biasing force method~\cite{abf}. 
These last two biased MD simulations improve drastically the sampling of rare events.
We also estimated the minimum potential energy path between the wells at 0~K by running geometry optimizations with an 
added harmonic restraint potential $V\lp\delta\rp=\frac{k}{2}\lp\delta-\delta_0\rp^2$ on the collective variable $\delta$. 
We effect the mapping of the potential energy profile by moving the center $\delta_0$ of the parabola.
We found consistent energy profiles using the different sampling techniques.
In \myfig{comp}, we provide such comparison where we include two additional sampling methods: 
the blue-moon ensemble technique~\cite{bme} and the nudge elastic band method~\cite{neb}.

\section{Results and discussion}
\label{discussion}

\begin{figure*}
\includegraphics[width=\textwidth]{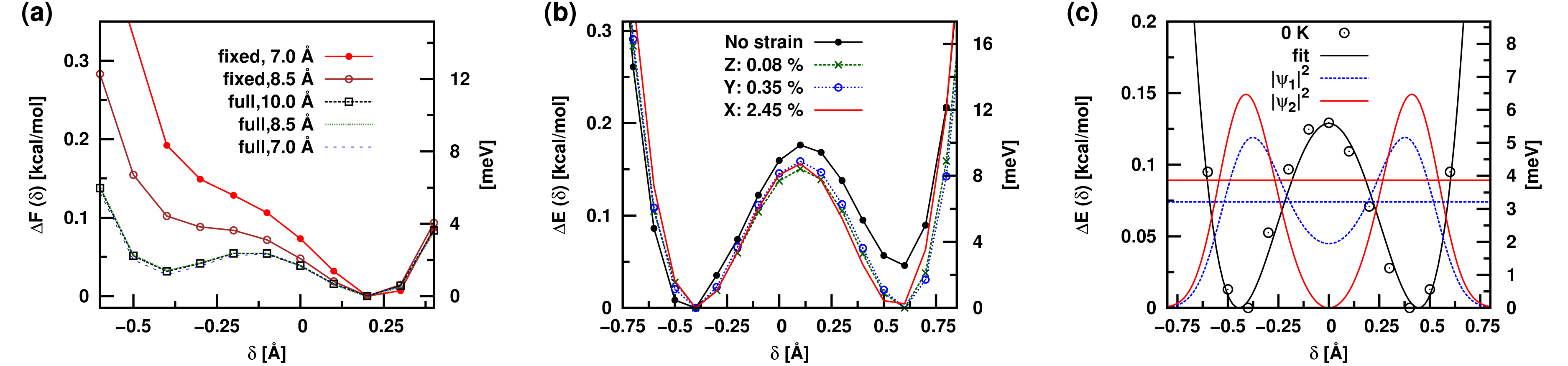}
\caption{\label{other}
(Color online) Calculated properties of the motif with the central aluminum atom surrounded by the distorted octahedron of oxygen atoms (\myfig{1stmotif}).
\bfa\ The effect of fixing the position of atoms beyond a given radius (in \AA) around the central Al atom on the minimum energy path between the minima (``fixed'', in legend).
The effect of changing the cutoff distance for short-ranged interactions is also shown (``full'', in legend) where no frozen atoms are imposed. The 1500-atoms system and force field I were used.
\bfb\ Tuning the asymmetry of the energy profile at 0~K by mechanical strain (given as \%) along X,Y, and Z axis. Results according to force field II~\cite{beck}.
\bfc\ Estimation of the tunneling splitting by solving the Schr\"odinger equation numerically using an effective mass $m_\mathrm{eff}=15$~a.m.u. and a fit of the profile at 0~K given
in \myfig{1stmotif}. Both eigenfunctions $|\psi_1|^2$ and $|\psi_2|^2$ were scaled down by a factor of 10 to fit in the graph.
Their associated eigenvalues are indicated by blue and red horizontal lines, respectively.
}
\end{figure*}

\myfig{1stmotif}\bfa\ shows some of the largest fluctuations in atomic positions found during 
a microcanonical (NVE) MD simulation of amorphous alumina at 5~K using 1500 atoms and force field I~\cite{matsui94}.
Using the asymmetric stretch $\delta=|\vec{r}_\mathrm{O2-Al}|-|\vec{r}_\mathrm{Al-O1}|$ as collective variable, 
we observe that for over 0.5~ns it displays the typical signature of a bistable ``rare event'' at that temperature, 
which corresponds to the transfer of a central Al atom along the line between two axial O atoms (\myfig{1stmotif}\bfb). 
At the transition state, the first coordination shell of the motif resembles a distorted octahedron with six O atoms around a central 
Al atom (typical coordination of corundum crystal) with elongated axial Al--O bonds, that are perpendicular to the equatorial plane (\myfig{1stmotif}\bfb). 
~\footnote{An animation of this motif is available at \url{http://nano-bio.ehu.es/users/alejandro}.}
A cavity defect and low temperatures are essential for this motif survival.
The same motif was also found with the force field II from the quenching of completely different melted alumina systems using different simulation parameters. 
This confirms that the structural motif in \myfig{1stmotif}\bfb\ is completely reproducible.

\myfig{1stmotif}\bfc\ shows the reconstructed free energy profiles at 0, 5, and 25~K computed using both force fields I~\cite{matsui94} and II~\cite{beck}.
The free energy difference between wells is about 2~meV in both cases and 
the largest energy barriers for the transition between minima are 4 and 12~meV for force fields I and II, respectively.
\myfig{comp} provides a further comparison of free energy profiles computed using several different sampling techniques and good agreement is found.
All our energy profiles are asymmetric because such configurations are entropically favored due to the inherent disorder of glasses.
Also, we do not have enough statistics to see these even ``rarer'' symmetric TLSs as they would demand extremely large system sizes and/or long simulation times.
Regarding the effect of temperature, it is well-known that the TLS only survives at low enough temperatures ($<100$~K).
\myfig{1stmotif}\bfc\ shows that raising the temperature increases the asymmetry of the double-well profile.
We note, however, that the effect is more evident for the force field II than for the I.
There is a non-trivial influence of the atoms surrounding the octahedron that will be discussed below.

\myfig{1stmotif}\bfb\ may give the wrong impression that the motif is rather localized and independent of its surroundings, 
however further tests show that this is not the case.
\myfig{other}\bfa\ shows that the environment around the motif has a non-trivial influence on its stability.
The motif involves tens of atoms rather than just one aluminum atom jumping between two axial oxygen atoms.
Here, we froze all atoms beyond a certain radius about the central aluminum atom and recomputed the energy profile (``fixed'', in legend) at 0~K along the collective variable $\delta$.
We see how the double-well profile deteriorates as we gradually shrink the sphere around central aluminum atom (that is, as more atoms are frozen).
We estimate the characteristic radius of the bistable motif to be $\sim$~8~\AA.
In \myfig{other}\bfa, we also investigated the impact of changing the cutoff distance in the short-range interactions (``full'', in legend) on the energy profiles and found little effect provided that a cutoff distance greater than 7~\AA\ is used. 
We further explored the finite-size effects on the energy profiles by re-optimizing a larger cell containing the TLS motif in \myfig{1stmotif}\bfb. 
A $2\times 2\times 2$ replication of our 1500-atoms system (12000 atoms in total) yields a similar energy profile at 0~K (see \myfig{2x2x2}) as the one shown in \myfig{1stmotif}\bfc. 
This result confirms that our motifs are independent beyond a certain radius and much less affected by the farthermost surroundings, in agreement with one of the basic tenets of the TLS model.

As mentioned, the energy profile of the identified motif is typically asymmetric due to the intrinsic disorder of the amorphous state, 
but it can be symmetrized by applying strain to the alumina sample. 
\myfig{other}\bfb\ demonstrates that the energy profile obtained by force field II on the 360 atoms system (\myfig{1stmotif}\bfc) becomes symmetric upon stretching by 0.1--2.5~\% depending on the strain direction. 
The coupling coefficient between strain and energy asymmetry is found to be $\gamma\sim$~0.2--1.0~eV 
for different strain directions, in good agreement with experiments~\cite{heuer94, phillips81, phillips87, grabovskij12}. 

\begin{figure*}
\includegraphics[width=\textwidth]{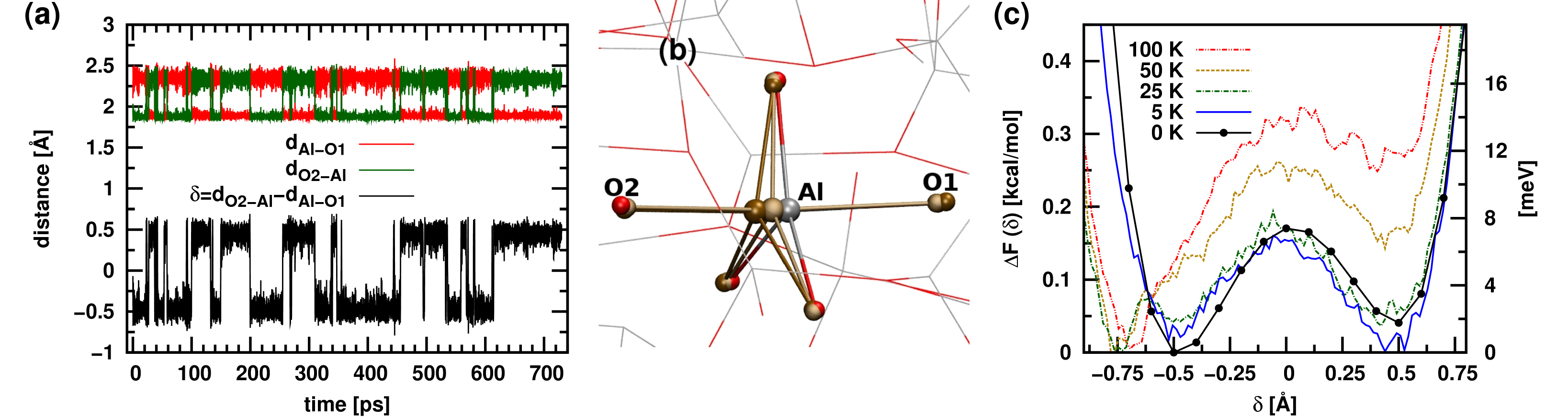}
\caption{\label{2ndmotif}
(Color online) Bistable motif with a central aluminum atom surrounded by a distorted trigonal bipyramid of oxygen atoms.
This Al atom oscillates between the two oxygen atoms O1 and O2.
\bfa\ Calculated time evolution of the relevant collective variables during a microcanonical MD simulation of at 25~K with force field I.
\bfb\ Superimposed snapshots of the optimized configurations corresponding to the initial (dark ochre color), transition (light color)
and final (aluminum atoms in grey, oxygen atoms in red) states at 0~K.
This motif resembles the one shown in \myfig{1stmotif} except that one of the equatorial oxygen atoms is missing.
\bfc\ Free energy profiles for the aluminum transfer along the collective variable $\delta=|\vec{r}_\mathrm{O2-Al}|-|\vec{r}_\mathrm{Al-O1}|$ at 0, 25, 50, and 100~K computed using force field I~\cite{matsui94}.
}
\end{figure*}

Another property of TLSs that can be measured is its direct coupling to external fields. 
Using the partial charges of force field II, we estimated the change in electric dipole moment $\delta p$ for motif in \myfig{1stmotif} from
$$\delta p=\left[\left(p_x^i-p_x^f\right)^2+\left(p_y^i-p_y^f \right)^2+\left(p_z^i-p_z^f\right)^2\right]^{1/2},$$
where $i$ and $f$ refer to our initial and final local minimum configurations, respectively, in the energy profile of \myfig{1stmotif}\bfc.
The transition between the two energy minima is associated with a change in electric dipole moment of $\delta p = 4.2$~D.
This value corresponds to a single electron charge moving by 0.9~\AA, in good agreement with experiments~\cite{martinis05, constantin07}. 

Nevertheless, the most important feature that determines whether a proposed structural motif can influence the low-temperature thermal properties of the amorphous material and its microwave spectroscopy is the tunnel splitting.  
To this end, we estimated the effective mass of the collective variable for the 0~K profile of \myfig{1stmotif}\bfc\ using \myeq{eq:meff}.
The calculated effective mass using both force fields I and II varies from 8 to 16 a.m.u. (\myfig{fig:meff}) along the minimum energy path. 
It is noted that the reduced mass for the asymmetric stretch of a system of 3 collinear atoms (O--Al--O) is $m_\mathrm{O} m_\mathrm{Al}/\lp 2m_\mathrm{O}+m_\mathrm{Al}\rp=7.3$~a.m.u.. 
This is another manifestation that not only one atom participates in the transition between the two energy minima.
Using the full coordinate dependence of the effective mass (\myfig{fig:meff}) we estimated a tunnel splitting of $\delta E = 0.3-0.7$~meV using instanton theory.
Taking an effective mass equal to $m_{\mathrm{eff}}\sim 15$~a.m.u., we also 
solved the one-dimensional Schr\"{o}dinger equation numerically for a symmetrized version of potential energy profile in \myfig{1stmotif}\bfc. 
The estimated value of 0.65~meV (\myfig{other}\bfc) is somewhat higher than the experimental data~\cite{phillips81,phillips87,lupascu09,grabovskij11,grabovskij12,simmonds04,martinis05,lisenfeld10a,lisenfeld10b,palomaki10}. 
However, smaller values of the tunnel splitting can be achieved for the motifs with higher barriers and larger well separations.
We note that reproducing such small experimental splittings is beyond the precision of current ordinary \ai\ calculations.

After inspecting \myfig{1stmotif}\bfb, it is reasonable to wonder if a similar motif is feasible but with only three oxygen atoms in the equatorial plane.
\myfig{2ndmotif}\bfb\ shows that this is indeed possible.
This new motif has a trigonal bipyramidal shape and was found using the force field I~\cite{matsui94}.
\myfig{2ndmotif}\bfa\ shows the typical time evolution of the relevant collective variables and displays a clear bistable behavior.
\myfig{2ndmotif}\bfc\ features a more symmetric profile and a higher energy barrier (8~meV) than its counterpart in \myfig{1stmotif} (4~meV).

\begin{figure}
\includegraphics[scale=0.3]{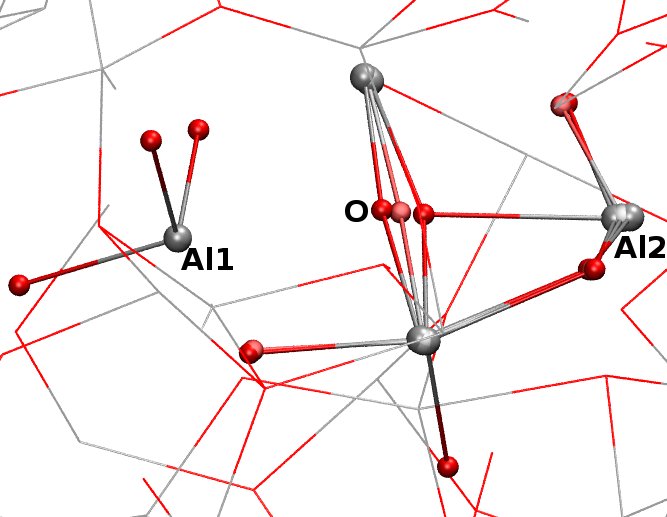}
\includegraphics[width=\columnwidth]{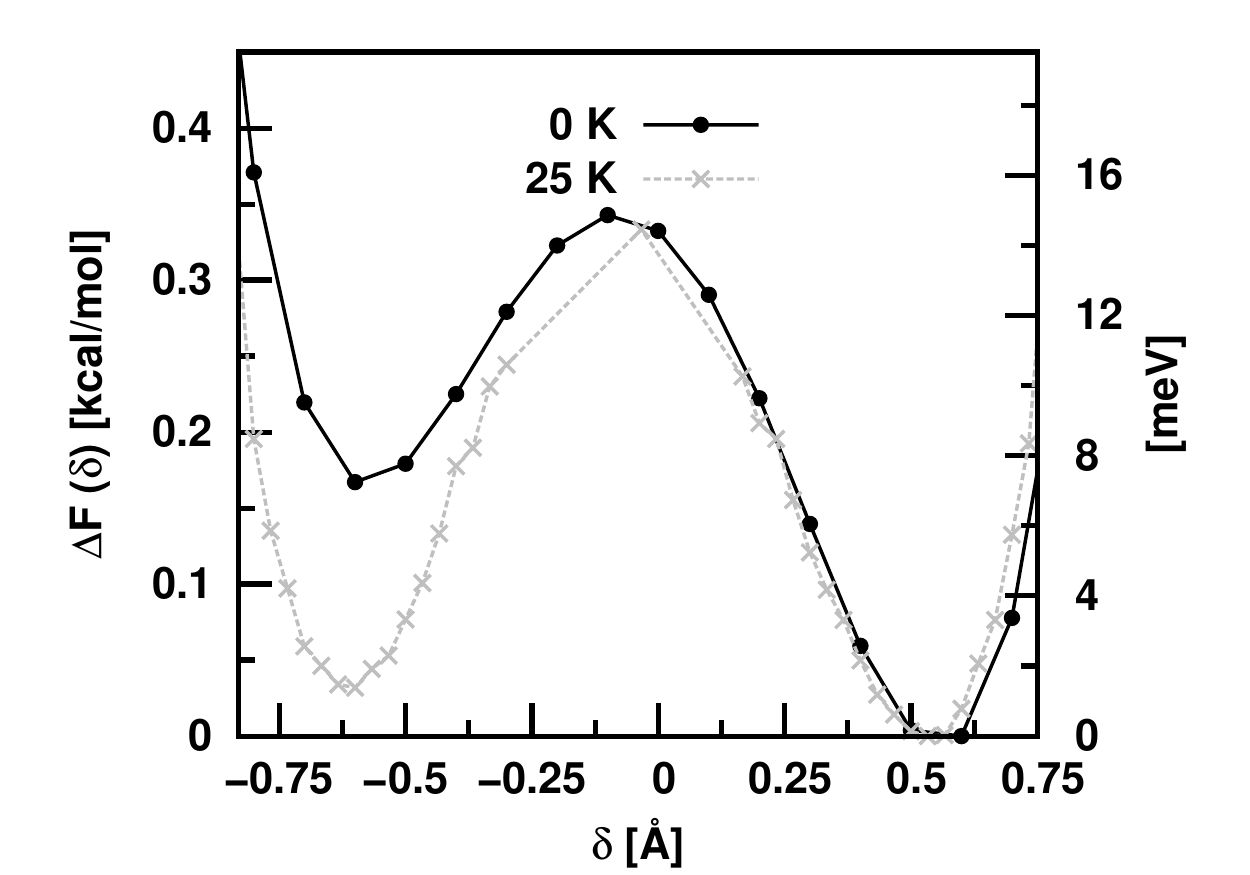}
\caption{
(Color online) Bistable motif with a mobile central oxygen atom surrounded by four aluminum atoms and oscillating along the line between two aluminum atoms (Al1 and Al2).
Top: Superimposed snapshots corresponding to the initial, transition, and final states for the motif at 50~K.
Aluminum and oxygen atoms are depicted as grey and red spheres, respectively.
Bottom: Free energy profiles computed along the asymmetric stretch $\delta=|\vec{r}_\mathrm{O}-\vec{r}_\mathrm{Al1}|-|\vec{r}_\mathrm{O}-\vec{r}_\mathrm{Al2}|$
at 0 and 50~K using force field II~\cite{beck}.
\label{3rdmotif}
}
\end{figure}

Are these all the possible structural TLS motifs in amorphous alumina?
We think that the answer to this question is negative. 
There exists an ensemble of different structures responsible for the low-temperature behavior and that we just found a few of them.
For instance, in \myfig{3rdmotif} we show another plausible TLS motif found for the 360-atoms system using the force field II~\cite{beck}.
This motif is characterized by a central tetra-coordinated oxygen atom (typical coordination in the crystal) 
jumping back and forth between two aluminum atoms, labelled as Al1 and Al2 in the \myfig{3rdmotif} (Top).
Although a mobile oxygen atom was also proposed by DuBois and coworkers~\cite{dubois13}, our motif geometry is very different to theirs. 
We remark that their motif was an \textit{ad-hoc} model and might not be stable in the real amorphous environment, 
while ours is the direct result of MD simulations that consider explicitly the surrounding lattice.
Although we have not found their motif in our calculations, it would be interesting to explore its existence in further atomistic MD simulations of amorphous alumina.
In \myfig{3rdmotif} (Bottom) we show the corresponding free energy profile at 0 and 25~K along the collective variable $\delta=|\vec{r}_\mathrm{O-Al1}|-|\vec{r}_\mathrm{O-Al2}|.$
The plot features profiles with an energy barrier of the order of 15~meV.
Interestingly, the temperature seems to have a symmetrizing effect on the profiles. 

The relative abundance of found TLS motifs in our amorphous alumina systems was rather small, of the order of 1 motif per 10000 atoms.
This value was a rough estimate from the (limited) statistics of our MD simulations, that is, the number of TLS found (7) per the total number of atoms simulated ($>$50000).
Assuming a uniform energy distribution (a basic assumption in the standard TLS model) we estimate the density of experimentally relevant TLSs (those with energy asymmetry below 0.1~meV)
as 1 TLS per 200000 atoms, approximately.
A more accurate value of the TLS density could be calculated with the systematic search algorithm proposed by Reinisch and Heuer~\cite{reinisch04,reinisch05}
that improves upon standard methods that may miss several TLS candidates.

\section{CONCLUSIONS}
\label{conclusions}

In conclusion, using extensive computer simulations, we show that bistable structures are naturally present in amorphous alumina at temperatures below 100~K. 
We found several structural motifs that form TLSs in alumina.
We also found that the most mobile atoms of the TLSs could be either oxygen or aluminum atoms.
We crudely estimate the density of experimentally relevant TLSs (those with energy asymmetry below 0.1~meV) to be 1 TLS per 200000 atoms. 
This value was estimated from the number of TLS found (7) per the total number of atoms simulated ($>$~50000), 
and from the standard assumption of the TLS model that the energy distribution of TLSs is uniform.
In particular, we identified a motif resembling a distorted octahedron where the transferring aluminum atom performs an umbrella-like motion between the two axial oxygens. 
The robustness of this motif was confirmed by using two different force fields, different MD simulation parameters, and by cooling different alumina melts. 
The properties of the corresponding TLSs are consistent with experimental observations in Refs.~\cite{grabovskij12}. 
Combination of our results with the microwave spectroscopy of TLSs~\cite{grabovskij12} 
opens a venue for the eventual understanding, and, possibly, utilizing TLSs in amorphous alumina. 
Our results suggest that similar motifs may exist in other amorphous materials with a local corundum structure (see \myfig{hem}). 
The extension to such systems as well as the effect of pressure on TLSs are interesting problems for the future.

\begin{acknowledgments}
We acknowledge financial support from the Marie Curie IIF (PIIF-GA-2012-326435 RespSpatDisp), 
the ERC Advanced Grant DYNamo (ERC-2010-AdG-267374),``ayuda para la Especializaci\'{o}n de Personal Investigador del Vicerrectorado de Investigaci\'{o}n UPV/EHU-2013'', Grupos Consolidados UPV/EHU del Gobierno Vasco (IT-578-13), and Ikerbasque.
This research was supported, in part, by a grant of computer time from the CUNY HPC under NSF Grants CNS-0855217, CNS-0958379 and ACI-1126113. 
We thank Alexey Ustinov and Grigorij Grabovskij for suggesting this problem and for useful discussions.
\end{acknowledgments}

\appendix

\section{Pair distribution functions and static structure factors}

\begin{figure*}
\includegraphics[width=\textwidth]{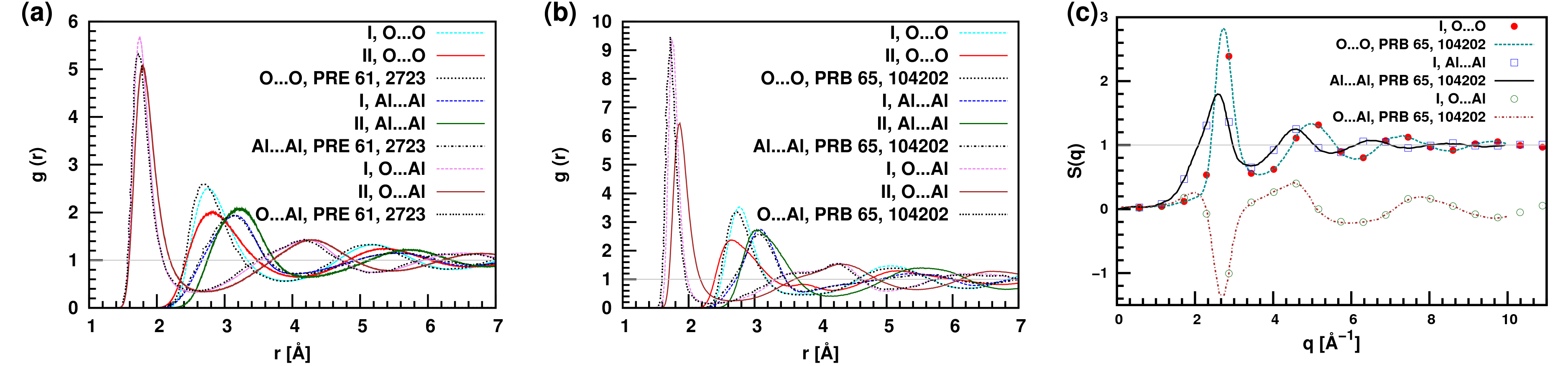}
\caption{
(Color online) Comparison between our computed equilibrium structural properties with previous calculations~\cite{gutierrez00,gutierrez02}.
\bfa\ Comparison of radial distribution functions for liquid alumina at 2200~K with force fields I and II with previous calculations.
\bfb\ Comparison of radial distribution functions for amorphous alumina at 650~K for force fields I and II versus others.
\bfc\ Partial static structure factors computed for amorphous alumina (1500 atoms) at 650~K using force field I versus previous calculations.
}
\label{rdf}
\end{figure*}

We have validated our liquid and amorphous alumina configurations by comparing the calculated pair distribution functions and partial static structure factors
to previous calculations~\cite{gutierrez00,gutierrez02}.
We provide such comparison in \myfig{rdf}\bfa\ and \myfig{rdf}\bfb, where we show the radial distributions functions for liquid alumina at 2200~K and its corresponding
amorphous phase at 650~K, respectively.
These temperatures were chosen to facilitate the comparison with previous computational data~\cite{gutierrez00,gutierrez02}.
We report our results for the 1500-atoms system with force field I and for the 360-atoms system with force field II.
\myfig{rdf}\bfc\ also shows a comparison between our computed partial static structure factors for the different pairs with previous calculations using force field I~\cite{gutierrez00}.
Our calculations using force field I are in excellent agreement with previous similar calculations, which validates our alumina configurations.

\begin{figure}
\includegraphics[width=\columnwidth]{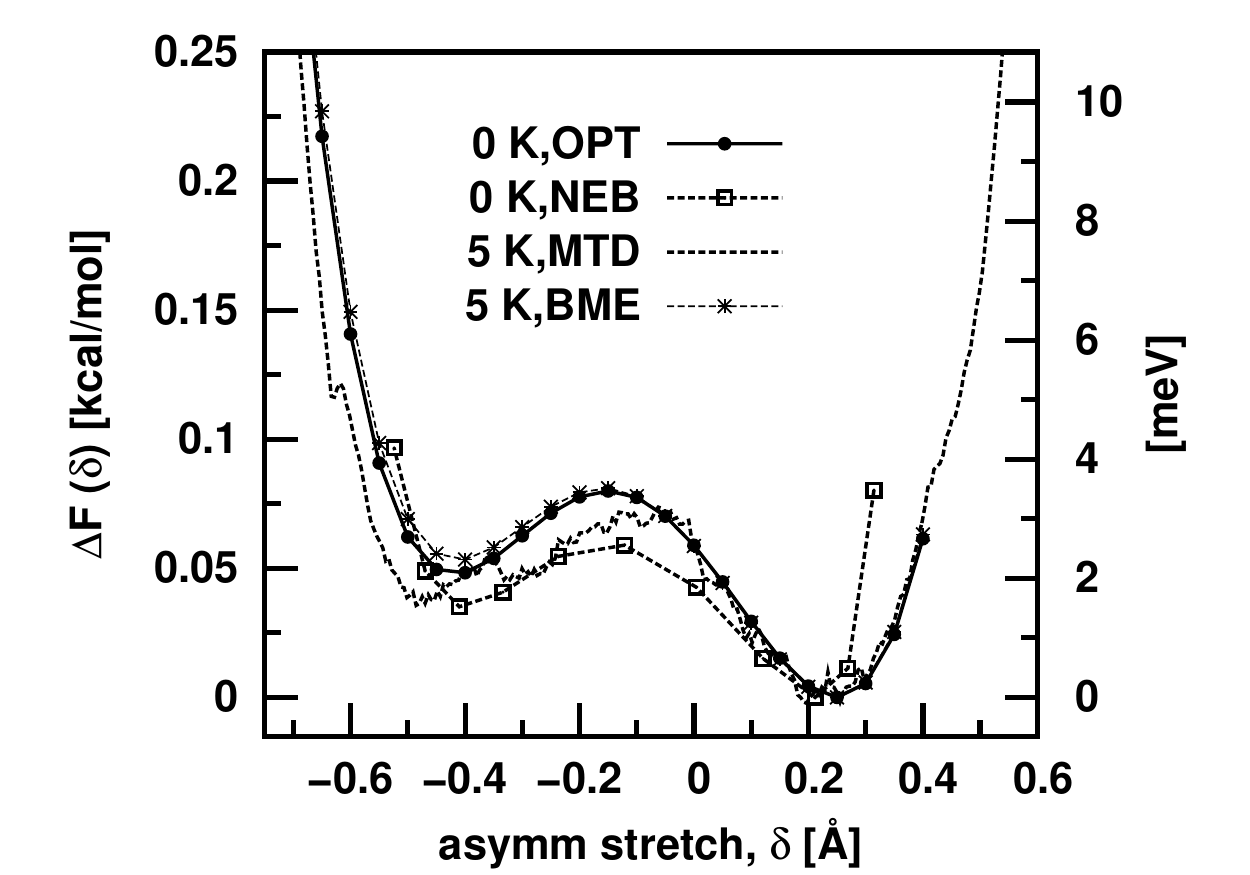}
\caption{
Comparison of free energy sampling methods for the motif in \myfig{1stmotif}\bfb\ (1500 atoms and force field I).
At 5~K, we compare metadynamics (MTD, in legend)~\cite{mtd} with the blue-moon ensemble (BME)~\cite{bme} method.
At 0~K, the minimum energy path was mapped with static restrained optimizations (OPT) and with the nudge elastic band (NEB) method~\cite{neb}.
}
\label{comp}
\end{figure}

\section{Comparison of different free energy sampling techniques}

Different sampling methods used to compute the energy profile in \myfig{1stmotif}\bfc\ are compared in \myfig{comp}.
The alumina system investigated was the one with 1500 atoms described by the force field I.
Good agreement is found for all methods.

\section{Finite-size effects on energy profiles}

We compared the energy profile at 0~K for the motif in \myfig{1stmotif}\bfb\ (1500 atoms and force field I)
with a $2\times 2\times 2$ replicated version of that system (a total of 12000 atoms)
using different cutoff radius for the pairwise interactions.
Good agreement is found for the different profiles.

\begin{figure}
\includegraphics[width=\columnwidth]{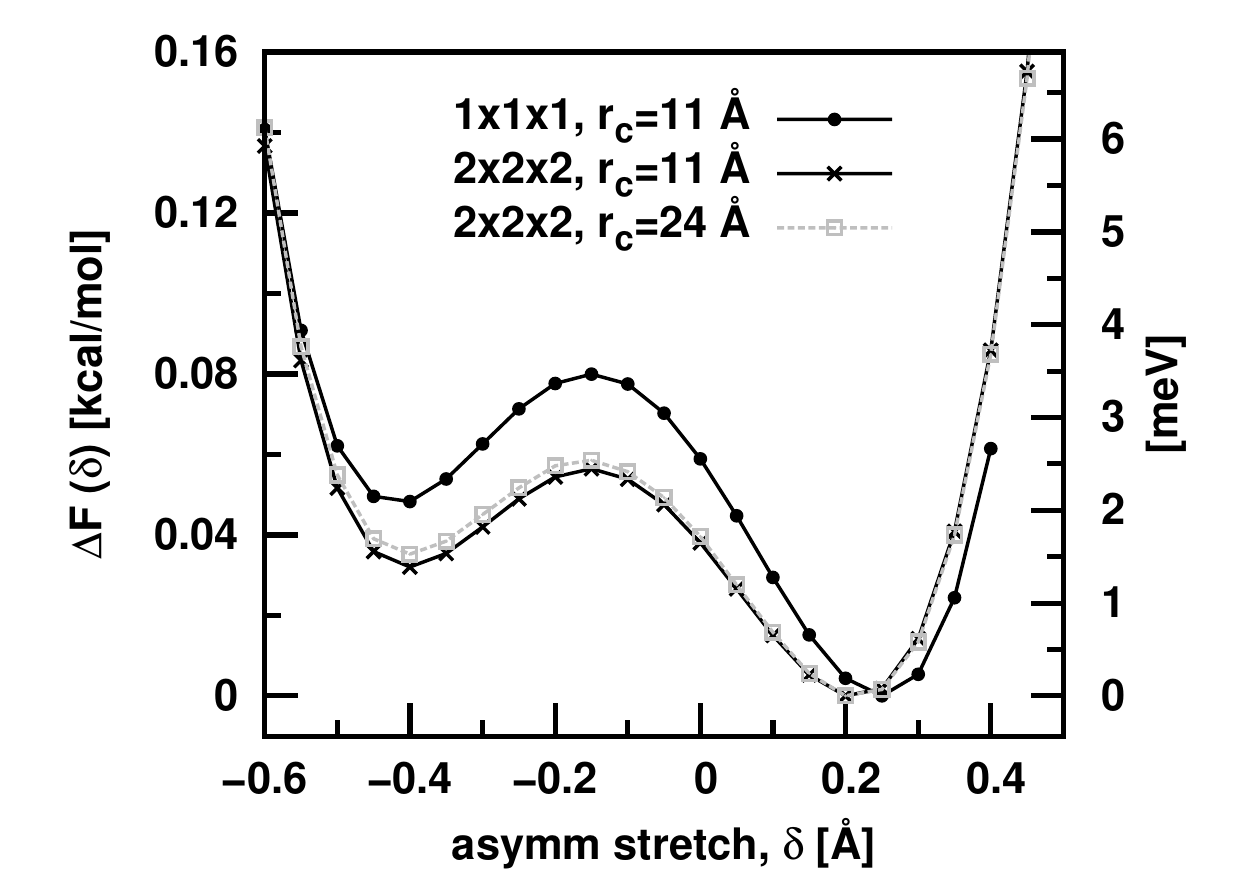}
\caption{
Comparison of energy profiles at 0~K for the motif of \myfig{1stmotif}\bfb\ with a $2\times 2\times 2$
replication of such system using different cutoff radius for the interactions.
}
\label{2x2x2}
\end{figure}

\section{Maximum size of avoided level crossing}

It is also possible to estimate the maximum size of the avoided level crossings, $S_{\mathrm{max}}$, expected in microwave spectra due to the qubit interaction with the TLS~\cite{martinis05, dubois13}

\bea
S_{\mathrm{max}}&=&2\frac{\delta p}{w h} \sqrt{\frac{E_{01}}{2C}},
\label{S}
\eea

\noindent where $h$ is the Planck's constant, $w$ is the width of the amorphous oxide barrier, $C$ is the capacitance, and $E_{01}=\delta E$ is the energy difference between the two lowest qubit levels.
Using typical parameters ($w=2$~nm, $C=1$~pF, $E_{01}=0.1$~meV), \myeq{S} gives a TLS-qubit coupling strength $S_{\mathrm{max}}\approx 60$~MHz, in good agreement with the experimental data ($\sim 100$~MHz)~\cite{simmonds04, martinis05,lupascu09,lisenfeld10a, lisenfeld10b, palomaki10,grabovskij11,grabovskij12}.

\section{Calculation of the effective mass and tunneling splitting}

To estimate the effective mass associated to the motif, we used three methods:
(1) We used the equipartition theorem on the collective variable $\delta$ 
\bea
m_\mathrm{eff}&=&k_\mathrm{B}\frac{\la T_{\delta}\ra}{\la\dot{\delta}^2\ra}.
\label{equipartition}
\eea 
Here, the angular brackets denote a thermal average over the canonical ensemble.
The variables $T_{\delta}$ and $\dot{\delta}$ are the temperature and the velocity of the collective variable, respectively.
The latter was evaluated analytically.
We are aware that application of equipartition theorem to non-ergodic systems such as amorphous alumina is questionable.
(2) We performed a Fourier transform of the time evolution of the collective variable $\delta\lp t\rp$ from a microcanonical MD trajectory to get characteristic frequencies, $\omega$, in the lower part in the spectrum. 
For the motif in \myfig{1stmotif}, we found a large vibrational peak at $\nu=\omega/2\pi\sim 4.12$~THz.
Then, from the known curvature of one of the wells ($\kappa=3.62$~J~m$^{-2}$), 
we estimated the effective mass according to the harmonic approximation: $m_\mathrm{eff}=\frac{\kappa}{\omega^2}$.
(3) The third and last method is based on classical mechanics.
The effective mass associated to a collective variable $\delta$ is given by
\bea
m_\mathrm{eff}\lp\delta\rp&=&\sum_{i=1}^N m_i \lc \lp\frac{\partial x_i}{\partial\delta}\rp^2 + \lp\frac{\partial y_i}{\partial\delta}\rp^2 + \lp\frac{\partial z_i}{\partial\delta}\rp^2\rc,\nonumber\\
\label{eq:meff}
\eea
where the sum is taken over all atoms $N$, each with Cartesian coordinates $x_i,y_i$ and $z_i$. 
The partial derivatives in \myeq{eq:meff} were computed numerically by finite differences.

\begin{figure}
\includegraphics[width=\columnwidth]{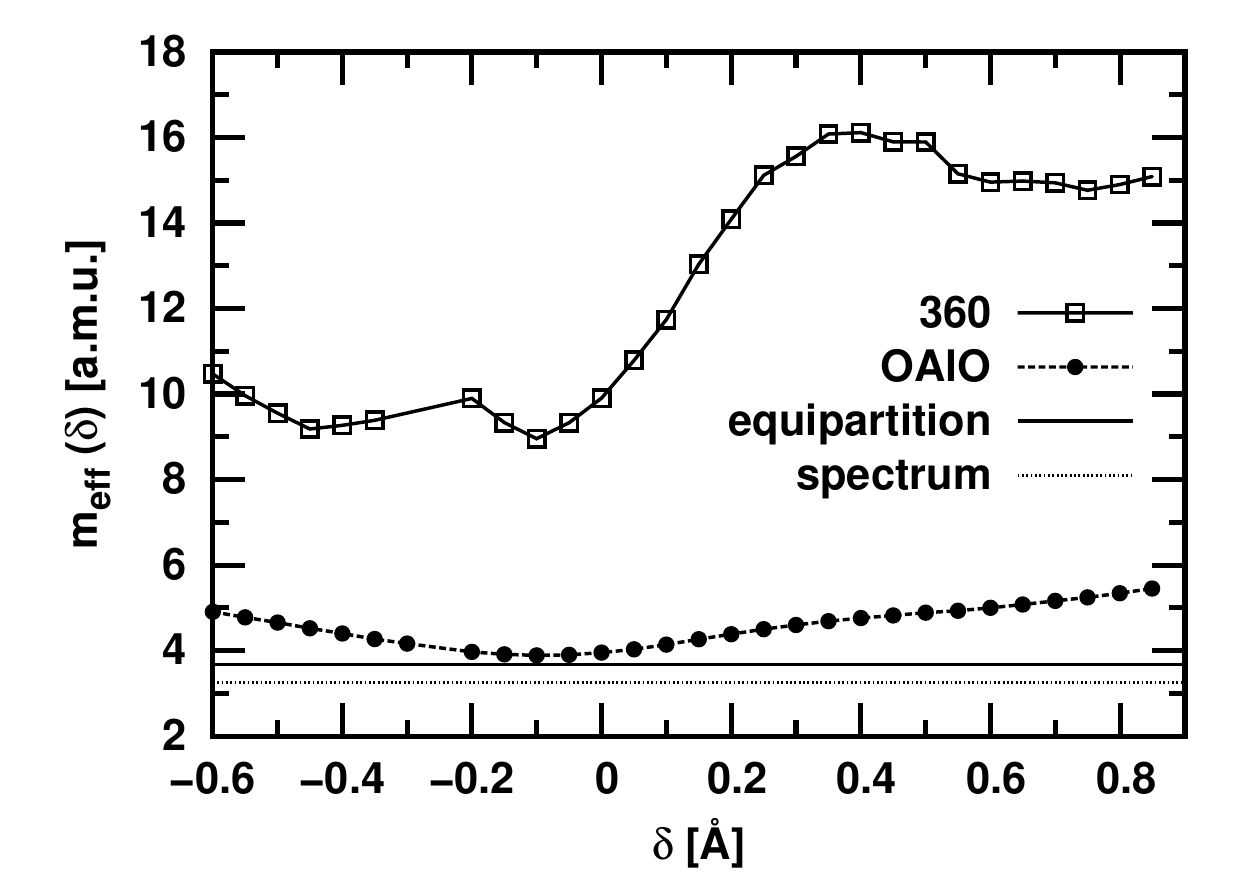}
\caption{
Estimation of the effective mass for the motif with the central aluminum atom surrounded by the distorted octahedron of oxygen atoms (\myfig{1stmotif}\bfb).
The effective mass of the collective variable $\delta=|\vec{r}_\mathrm{Al-O_2}|-|\vec{r}_\mathrm{Al-O_1}|$ was estimated using several techniques:
The equipartition method given by \myeq{equipartition}, the spectral method, and \myeq{eq:meff}. 
The two missing points at about $\delta=-0.3$~\AA\ were artifacts due to a ``kink'' in the minimum energy path and were eliminated from this graph.
}
\label{fig:meff}
\end{figure}

With these three methods, we estimated the effective mass for the motif shown in \myfig{1stmotif}\bfb.
This motif was found using the force field II on the 360-atoms system.
The results are in \myfig{fig:meff}.
Using only the coordinates of main three atoms (labelled as O1, Al, O2 in \myfig{1stmotif}\bfb) in \myeq{eq:meff} yields $m_\mathrm{eff}\approx 4$~a.m.u., 
which is consistent with the other two \textit{local} methods (equipartition and spectrum, in legend). 
However, this value seems to underestimate the truly effective mass ($m_\mathrm{eff}>9$~a.m.u.) 
because is missing the contribution from the rest, which is obtained using a summation over all atoms in the system (360, in legend).
We note that for the \textit{local} methods, $m_\mathrm{eff}$ is weakly dependent on the collective variable $\delta$, 
whereas if we consider all 360 atoms in \myeq{eq:meff}, the dependency becomes more pronounced.
On the left well (negative $\delta$), the effective mass for the full system was $m_\mathrm{eff}$=9~a.m.u., which is consistent with the expected reduced 
mass for a Al--O pair (10.04~a.m.u.), and increases as $\delta$ increases.
This is due to the fact that as we move along the reaction coordinate $\delta$, the differences between adjacent snapshots progressively increases. 

Based on the obtained potential energy profiles at zero temperature,
the characteristic barrier height $V_0$ of the TLS can be calculated by instanton theory~\cite{garg00}.
Due to the heavy atoms involved, we can safely assume that for qualitative estimates,
the instanton path is very similar to the calculated classical minimum energy path.
Therefore, the collective variable corresponding to motion along this instanton path
can be chosen the same as the generalized coordinate $q$ describing the activated transition between the two potential energy wells.
Then, the optimized action integral reads

\bea
\label{S0}
S_0&=&\int_{q_1}^{q_2}{\sqrt{2V\lp q\rp m_{\mathrm{eff}} \lp q \rp}\;\mathrm{d}q},
\eea
\noindent where the integration limits correspond to the turning points of the classical trajectory and $m_{\mathrm{eff}}(q)$ is the effective mass.
The tunnel splitting is given by $\delta E=\mathcal{F}\mathrm{exp}(-S_0/{\hbar})$,
where the ``fluctuation'' factor $\mathcal{F} $ is of the order of the vibrational energy $\hbar\Omega$ in the potential energy wells.
In general, neither the potential $V(q)$ nor the effective mass $m_{\mathrm{eff}}(q)$ in \myeq{S0} are constant along the trajectory.
However, we can approximate $S_0$ by using $V(q)\sim V_0$ and taking the effective mass equal to the mass of an aluminum atom, $m_{\mathrm{eff}}(q)\approx 27$~a.m.u..
Using the interwell distance $d=0.5$~\AA\space and a typical vibration frequency $\Omega=10^{13}\mathrm{s}^{-1}$,
we estimate that for an experimental splitting $\delta E=0.1$~meV, the characteristic barrier should be about $V_0\approx 8$~meV for alumina.
Thus, a temperature below 100~K should be enough to discern between thermally activated classical transitions between minima of TLSs and tunneling.

Once the effective mass of the collective variable was estimated, we fitted a symmetrized version of the energy profile at 0~K in \myfig{1stmotif}\bfb to a double-well like polynomial $V\lp\delta\rp=D\lp\delta^2-\delta_0^2\rp^2$.
The fitted values were $D=3.28$~kcal/(mol~\AA$^4$) and $\delta_0=0.45$~\AA.
We then solved the one-dimensional Schr\"odinger equation numerically using the code {\tt OCTOPUS}~\cite{octopus} to get the tunneling splitting between the ground and first excited states for a tunneling particle of mass m$_\mathrm{eff}\sim 15$~a.m.u..
In the {\tt OCTOPUS} calculation, we used an ultrafine grid spacing (0.01~a.u.) on a one-dimensional box of length 16~a.u..

\section{Other corundum materials: amorphous hematite}

Here, we address the question of whether the motifs found in amorphous alumina could also exists in the glassy state of other corundum-like materials.
To this end, we performed MD calculations of amorphous hematite (Fe$_2$O$_3$) at low temperatures.
The amorphous configuration was generated from the common crystalline phase $\alpha$-hematite at its experimental lattice parameters ($a=b=5.038$~\AA, $c=13.772$~\AA, $\alpha=\beta=90^\circ,\gamma=120^\circ$ and density=5.26~g~cm$^{-3}$) following the same melting-quenching protocol used for alumina.
An amorphous configuration of 1500 atoms (300 Fe$_2$O$_3$ units) was equilibrated in a orthorhombic cell with final volume $25.2762\times 25.2762\times 27.5973$~\AA$^3$, giving a mass density of 4.517~g~cm$^{-3}$ at zero external pressure. 
\begin{figure}
\includegraphics[width=0.45\columnwidth]{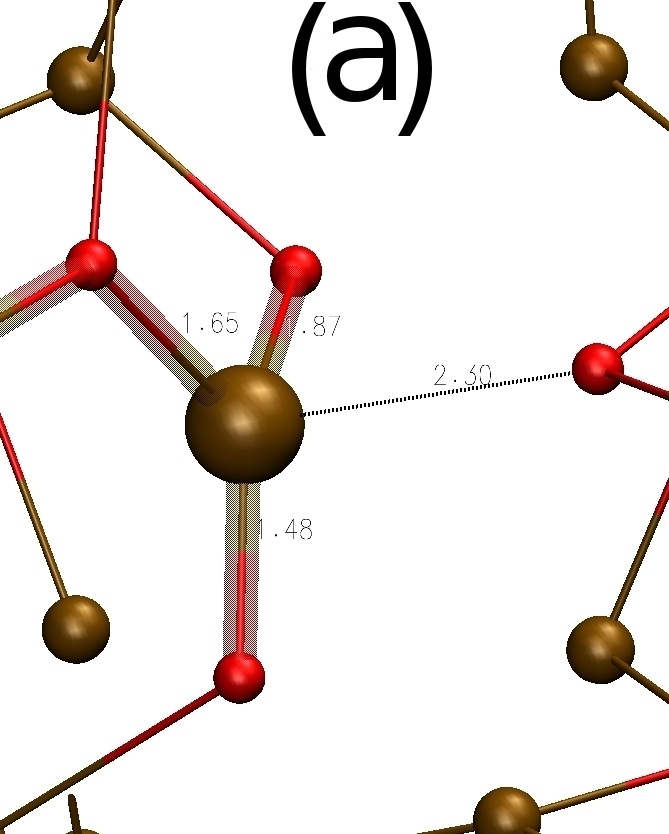}
\includegraphics[width=0.45\columnwidth]{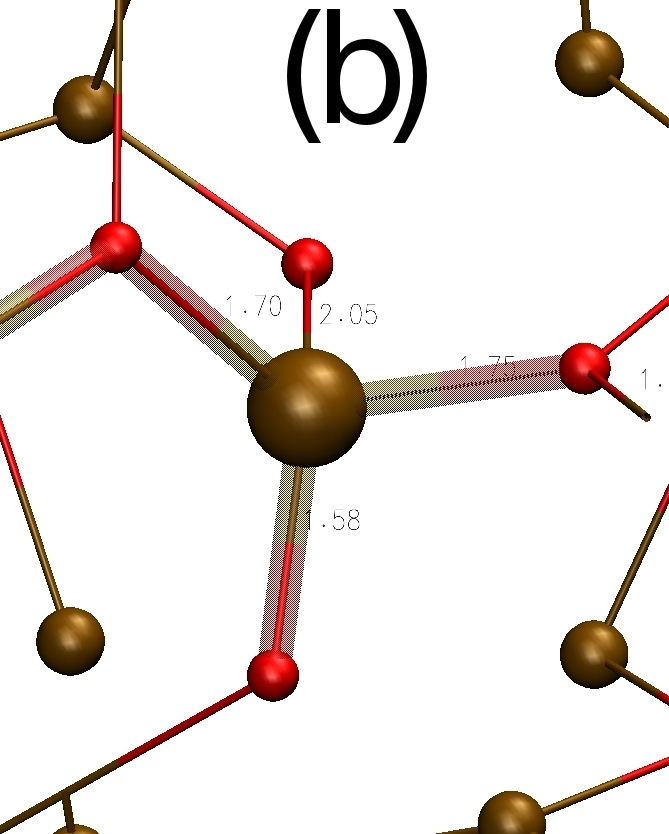}
\includegraphics[width=\columnwidth]{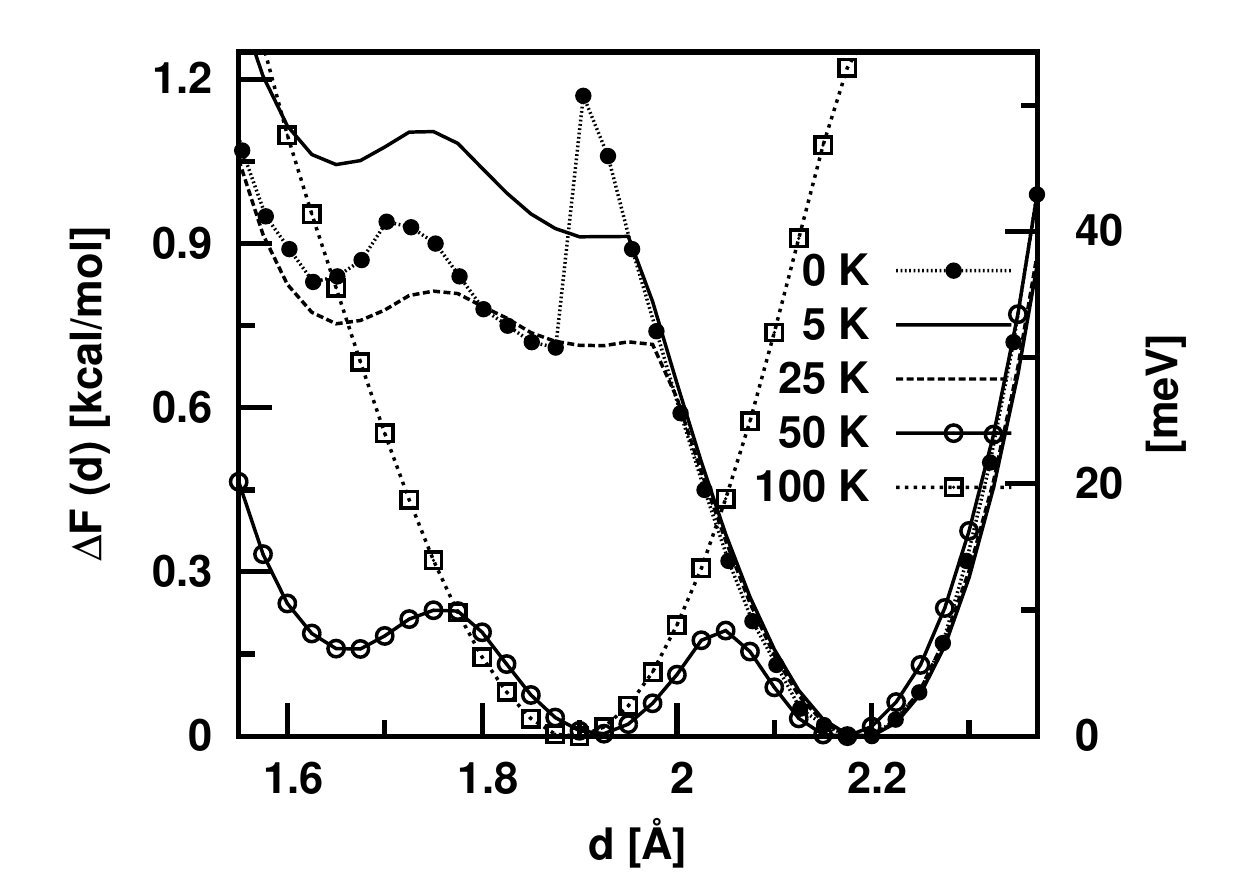}
\caption{
(Color online) A multilevel system in amorphous hematite (Fe$_2$O$_3$) found by our classical MD simulations at low temperatures.
Top: Superimposed snapshots corresponding to initial \bfa\ and final \bfb\ configurations in the energy landscape.
Interatomic distances are shown in \AA.
Iron and oxygen atoms are depicted as ochre and red spheres, respectively.
Bottom: Computed free energy profiles along the collective variable $d=|\vec{r}_\mathrm{Fe}-\vec{r}_\mathrm{O}|$ at several temperatures (in Kelvin).
\label{hem}
}
\end{figure}
For the MD calculations on Fe$_2$O$_3$, we used the reactive force field {\tt REAX}~\cite{reax-hematite} as implemented in the {\tt LAMMPS} code~\cite{lammps} together with its {\tt colvars} module~\cite{fiorin13} for the free energy calculations.
We used a time step of 0.2~fs and performed the charge equilibration~\cite{goddard91,nakano97,aktulga12} at every MD step up to a tolerance of $10^{-6}$~eV.
After several nanoseconds of MD simulation, we found only the motif shown in \myfig{hem}~(Top) that
resembles the transformation between a trigonal-pyramidal to a tetrahedral geometry.
This transformation, which is a change of coordination number (from 3 to 4) of the central iron atom,
can be followed by a single reaction coordinate consisting of the distance between the involved iron and oxygen atoms.
\myfig{hem}~(Bottom) shows the computed free energy profiles of this process at several temperatures.
This figure features a more complex multilevel landscape than the previous rearrangements found for amorphous alumina.
Interestingly, lowering the temperature bias the profile towards a single well corresponding to the trigonal-pyramidal configuration.
Thus, it is not clear whether this structural motif behaves as a TLS and remains to see whether amorphous hematite actually has TLSs.

\bibliography{refs3}

\end{document}

%% file: ff.tex
\begin{table}
\caption{Parameters for the force field I:~\cite{matsui94} 
$V_\mathrm{I}\left(r_{ij}\right)=A_{ij}\exp{\left(-r_{ij}/\rho_{ij}\right)}-\frac{C_{ij}}{r^6_{ij}}+\frac{q_iq_j}{r_{ij}},$ 
where $q_\mathrm{Al}=1.4175|e|$ and $q_\mathrm{O}=-0.9450|e|$ are the partial charges for aluminum and oxygen atoms, respectively.} \label{ffI}
\begin{ruledtabular} 
\begin{tabular}{lccc}
$i-j$ pair & $A_{ij}$, eV & $\rho_{ij}$, \AA & $C_{ij}$, eV~\AA$^6$\\
\hline
Al-Al & 31\,570\,911.694 & 0.068 & 14.051 \\
Al-O  & 28\,476.897 & 0.172 & 34.578 \\
O-O   & 6\,462.668 & 0.276 & 85.092 \\
\end{tabular}
\end{ruledtabular}
\end{table}
\begin{table}
\caption{Parameters for the force field II:~\cite{beck}
$V_\mathrm{II}\left(r_{ij}\right)=D_{ij}\left\{\exp{\left[\gamma_{ij}\left(1-\frac{r_{ij}}{\rho_{ij}}\right)\right]} -2 \exp{\left[\frac{\gamma_{ij}}{2}\left(1-\frac{r_{ij}}{\rho_{ij}}\right)\right]}\right\}+\frac{q_iq_j}{r_{ij}},$ where $q_\mathrm{Al}=1.244690|e|$ and $q_\mathrm{O}=-0.829793|e|$ are the partial charges for aluminum and oxygen atoms, respectively.} \label{ffII}
\begin{ruledtabular} 
\begin{tabular}{lccc}
$i-j$ pair & $D_{ij}$, eV & $\rho_{ij}$, \AA & $\gamma_{ij}$\\
\hline
Al-Al & 0.002164 & 5.517666 & 10.855181 \\
Al-O  & 1.000003 & 1.880153 & 7.617923 \\
O-O   & 0.000018 & 6.609171 & 16.719817 \\
\end{tabular}
\end{ruledtabular}
\end{table}